\documentclass[letterpaper]{jpconf}
\usepackage{graphicx}
\begin{document}
\title{Event characterization in (very) asymmetric collisions}

\author{Gabor David}

\address{Brookhaven National Laboratory, Upton, NY 11973, USA}

\ead{david@bnl.gov}

\begin{abstract}
Event-by-event reconstruction of the collision geometry 
using some incarnation of the Glauber-model is a widely accepted
method in studying heavy ion collisions.  While there is no known
problem with the procedure when applied to the collision of two large
ions, we will argue that in very asymmetric collisions, like $p(d)$+A
with at least one hard scattering process occuring the event geometry
deduced with the simple Glauber-model may be biased.
\end{abstract}

\section{Introduction}

Recent results from very asymmetric collisions at RHIC and LHC
(so far $p(d)$+A) provided several surprises, questioning the
decade-old assumption that such reactions probe the 
{\it cold nuclear matter} 
(CNM) effects only, and as such, serve as reference,
quasi-calibration of {\it no medium} effects,  which can then be
compared  to the A+A collisions, where a hot, dense medium is formed.  
This prevailing view was, however, fundamentally shaken in 2013.
At relatively low transverse momenta, where such phenomena are
typically associated with genuine hydrodynamic flow due to a strongly
interacting medium, long-range quadrupole azimuthal correlations have 
been observed at LHC in $p$+Pb
collisions~\cite{cms2013,alice2013,atlas2013} 
and subsequently in $d$+Au collisions at RHIC~\cite{ppg152}.
At the high end of the available $p_T$ range preliminary
results~\cite{baldo2013} indicated a significant change of the nuclear
modification factor from peripheral to central $d$+Au collisions, both
for single particles ($\pi^0$, $\eta$) and reconstructed jets.  This
result was clearly in tension with earlier findings and theoretical
expectations.  

In general terms the nuclear modification factor for particle species
$X$ and nuclei $A,B$ is defined as
$$ R^X_{BA} = \frac{dN^X_{BA}/dp_Tdy}{<N_{coll}>dN^X_{pp}/dp_Tdy}$$
\noindent
in case of $p(d)$+A collisions $B$ is simply just one proton or a
deuteron.  The crucial quantity is $N_{coll}$, the number of binary
nucleon-nucleon collisions in the overlap region of the two species
involved.  At the very least, $N_{coll}$ depends on the nuclear
geometry: the impact parameter of the collision, the density
fluctuations of the nucleus, the (energy dependent!) cross section of
nucleon-nucleon collisions.  In addition other kinematic and dynamic
factors may play a role.  In this paper we will raise the issue
whether the methods to determine collision geometry and subsequent
derivation of $N_{coll}$, worked out and
functioning well for collisions of large nuclei, is unquestionably
valid for very asymmetric ($p/d$+A) systems as well, or there are some
legitimate concerns?  Will the presence of a very hard scattering
change the overall event characterization?  Can such a change - if it
exists - be verified experimentally?

\section{Centrality in theory and in the experiment - large A+A collisions}

Detailed study of the properties of the sQGP relies
heavily on event-by-event classification of the collisions according
to the (implied) collision geometry.  Theorists need to know the
impact parameter $b$, whose magnitude defines {\it collision centrality}
in the purest sense, in order to calculate the nuclear overlap
$T_{AB}$ which then, combined with the nucleon-nucleon cross section
lead to quantities like the number of participant nucleons
($N_{part}$), the number of binary nucleon-nucleon collisions
($N_{coll}$), as well as the spatial distribution of participating
nucleons and quantities derived from it like eccentricity. 

Alternately, one can sidestep the question what $b$ is and define the
collision centrality with the number of nucleons participating in the
collision ($N_{part}$).  Just as $b$, $N_{part}$ cannot be directly
determined in the experiment, it is derived from some simple
{\it global  observable} like charged particle multiplicity ($N_{ch}$ )
or transverse energy ($E_T$) in a specific pseudorapidity region.
Since we are not discussing any particular experiment, we simply call
the detector(s) covering this region and serving both to trigger
{\it minimum bias} events as well as establish collision centrality,
Trigger Centrality Detector(s) (TCD). 
It should be noted that the TCD is usually located
close to beam rapidity and far from the rapidity region where the
actual signal is measured.  The global observables,
including the signal in TCD, are correlated with the directly unaccessible
$N_{part}$ - the nature of the correlation is discussed extensively
in~\cite{ppg100} along with a historic overview how our understanding
of the underlying processes evolved in the past decades.  In case of
large colliding nuclei and in an average event (no particles/jets above
a few GeV/$c$ observed) the $N_{part}$ vs $N^{TCD}_{ch}$ correlation is 
quite narrow, so $N^{TCD}_{ch}$ is a reasonable proxy for the unmeasured
$N_{part}$ (of $b$, for that matter).  The correspondence is usually
established with a Glauber Monte Carlo or with some
event-generator~\cite{annrev2007,ppg160}, by convolving the $N_{part}$
distribution with the (known or assumed) single-collision soft
production, and comparing it to the measured $N^{TCD}_{ch}$
distribution.  Once a good match between model and experiment is 
achieved, the measured distribution is divided up to percentiles, 
and the corresponding $<N_{part}>$, $<N_{coll}>$ established from the
model~\cite{annrev2007}.

A crucial fact in large-on-large A+B ion collisions is that the 
number of participating 
nucleons\footnote{We are aware that a recent paper~\cite{ppg100}
  shifts the emphasis from participating nucleons to quark
  participants, as the scaling variable for $N_{ch}$, but this doesn't
  change the essence of our arguments.
}
 from {\it both} nuclei 
($N^A_{part}, N^B_{part}$) is large in all but the most extremely 
peripheral collisions.  Combined with the observation that soft
production per participant pair, the basis to determine centrality, 
fluctuates, collisions with $N^A_{part}, N^B_{part}$ are practically
indistinguishable from collisions with $N^A_{part}, N^B_{part}-1$.  In
other words, even if some process would reduce (or outright eliminate)
the contribution from one nucleon to $N_{ch}$ production, this would
be experimentally undetectable.  We believe that this is exactly the
reason why the Glauber-model works well on large-on-large systems.
The same is not necessarily true when very asymmetric systems
collide, like $p(d)$ with a large nucleus A.

\section{The case of $pp$ and $p(d)+A$ collisions}

\begin{figure}
\centering{
\includegraphics[width=0.9\linewidth]{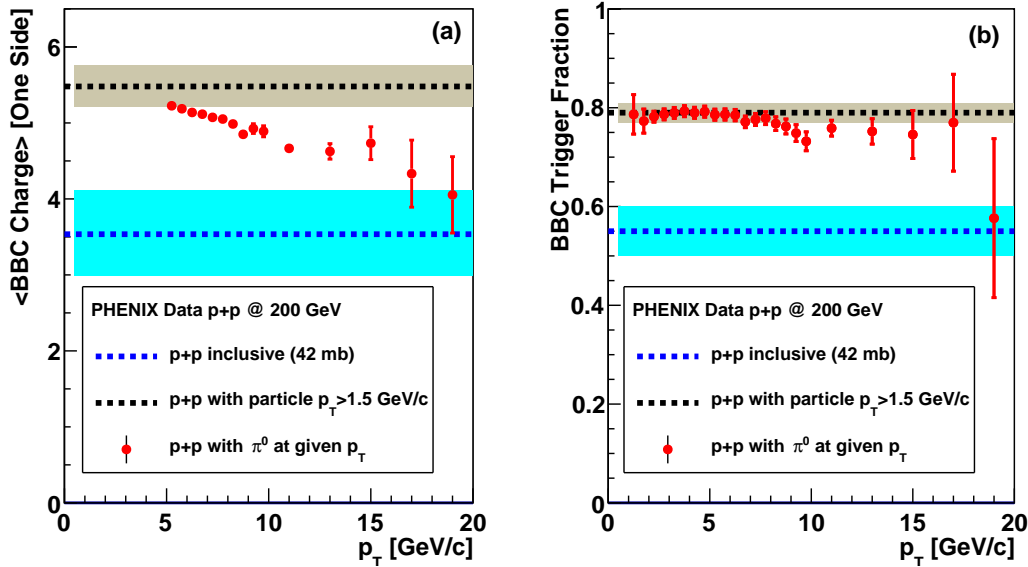}}
\caption{\label{fig:ppbiasdata}
  Left (a): $N^{TCD}_{ch}$ at $-3.9<\eta<-3.1$ vs the highest $p_T$ observed
  in a single particle at $|\eta|<0.35$ in $pp$
  collisions~\cite{ppg160}.  The two dashed lines are the mean charge
  for events taken with minimum bias trigger (lower, blue) and requiring
  at least one particle with $p_T>1.5$GeV/$c$ at midrapidity.
  Right (b): Trigger efficiency (probability of the coincidence of at
  least one particle at both $-3.9<\eta<-3.1$ and $3.1<\eta<3.9$) for
  minimum bias events (lower, blue line), events with at least one
  particle with $p_T>1.5$ at midrapidity (upper, black line), and the
  dependence on the highest $p_T$ particle observed at midrapidity.
}
\end{figure}

It is interesting that in the only experimentally verifiable
case - namely $pp$-collisions - the situation is somewhat more
complicated.  There are two well-identified issues with $N^{TCD}_{ch}$
production near beam rapidity: the {\it trigger bias} and the 
bias on $N^{TCD}_{ch}$ if a high $p_T$ particle is observed far 
from beam rapidity (say, at $y=0$).  The two effects are
demonstrated in a very compact way in Fig.~\ref{fig:ppbiasdata} 
taken from~\cite{ppg160}.
To trigger an event requires either a coincidence of at least one hit
in each of two TCD detectors (up- and downstream), called minimum bias
trigger, or substantial activity (at least one particle with
$p_T>1.5$) at central rapidity.  Note that the later trigger is
significantly more efficient, i.e. once there is strong activity at
mid-rapidity, {\it soft} production forward and backward is more likely,
too.  On the other hand, if the activity at midrapidity is {\it too}
strong (rising maximum $p_T$), the trigger efficiency drops a few
percent, i.e. some of these high $p_T$ events are lost and the loss
has to be corrected for.

The left panel (a) is both more dramatic and more relevant for the
issue at hand, namely, centrality determination in very asymmetric
collisions.  It shows the mean value of $N^{TCD}_{ch}$ on one side for
minimum bias triggers (blue dashed line), for $p_T>1.5$ at midrapidity
(black dashed line) and as a function of the maximum $p_T$ observed at
midrapidity (red points).  The exact reason of the relatively fast 
depletion of $N^{TCD}_{ch}$ with increasing midrapidity $p_T$ is not 
clear, and herein lies the potential problem in determining centrality
in $p(d)$+A collisions from $N^{TCD}_{ch}$.  Obviously in the
asymptotic limit - two partons, both carrying $x\approx1$ fraction of
momentum and scattering with the maximum possible $q^2$ at mid-rapidity - 
$N^{TCD}_{ch}$ goes to zero, but the probability of this happening is
vanishingly small.  Still it is useful to keep in mind because at 
{\it some} $p_T$ this {\it kinematic} effect will start playing a role,
even if we don't know (in a model-independent, experimentally
verifiable way) where.  It is, however, unclear whether the
kinematic effect is sufficent to explain the entire drop
seen in Fig.~\ref{fig:ppbiasdata}.

Let's turn now to $p(d)$+A collisions.  Based on the success of the
Glauber Monte Carlo in large A+A collisions it is tempting to apply
the same method to derive centrality, and in fact, this is what has
been done early on.  In case of $d$+Au collisions there is an added
complication from the large size of the deuteron - the two nucleons
can be as far apart as 7-8\,fm, the radius of the entire Au nucleus -
but it also has an advantage: collisions in which only a proton or
neutron interacted can be tagged, providing important
cross-checks~\cite{ppg160}.  There is also a gradual shape-change and
shift in rapidity of the $N_{ch}$ distributions with centrality,
measured over a wide rapidity range both at RHIC~\cite{phobos2005} and 
at LHC~\cite{atlas2013-096,atlas2013-105}.  This rapidity shift has been
predicted (BGK triangle~\cite{gyulassy2005}) 
and it doesn't prevent us from
reliably re-calibrating centrality for the {\it average} event.
To emphasize the word {\it average}  
is more than pedantry: a comprehensive review of centrality
determination~\cite{annrev2007} explicitely states that {\it ``In heavy ion
collisions, we manipulate the fact that the majority of the initial
state nucleon-nucleon collisions will be analogous to minimum bias
$p$+$p$ collisions with a small perturbation from much rarer hard
interactions.''}   In fact, the most authentic source, Prof. Glauber
himself cautions the reader in his famous lecture,
p.340 in~\cite{glauber1959}:
{\it ``...limitations... the approximate wave function (74) is only
adequate for the treatment of small-angle scattering.  It does not
contain, in general, a correct estimate of the Fourier amplitudes
corresponding to large momentum transfer.''}  This exactly is the
justification why in the Glauber model - widely used to establish
centrality - all nucleon-nucleon collisions are considered to be
independent, occuring with the same cross-section and leading to
similar soft particle production, irrespective of their 
``history''\footnote{I.e. previous collisions and their ``violence''.
  We are aware that in a tightly coupled quantum-mechanical system the
  word ``history'' may be out of place - but that's exactly how the
  widely used Glauber model operates.
}.

What happens in {\it non-average} events, i.e. the ones where at least one
very hard scattering occurs?  As we've seen in the $pp$ case, the TCD
multiplicity - the basic building block of determining centrality -
changes substantially (see Fig.~\ref{fig:ppbiasdata}).  On the other
hand in the usual procedure the measured $N^{TCD}_{ch}$ is compared to
an $N_{part}$-fold convolution of the detector response
to an {\it average} nucleon-nucleon scattering.  While this is 
{\it  not} correct, judging from the experimentally verifiable $pp$
case, as long as both $N^A_{part}$ and $N^B_{part}$ are large, 
the mistake in assigning a centrality class to the event is minute.
However, this is {\it not} true in $p(d)$+A collisions, since there
are only one or two nucleons on one side.  Once a hard scattering
occured, this nucleon was necessarily part of it.  In fact,
in~\cite{ppg160} the authors dealt with the problem by introducing
$p_T$-dependent centrality bias factors, re-calculating the apparent
centrality by folding $N_{coll}-1$-times the {\it normal}
nucleon-nucleon response and once the reduced one.  The virtue of this
approach is that it changes the generally accepted, commonly used
experimental method in just one, clearly defined step.  The underlying
assumption, while debatable, is crisp: even if {\it the} projectile
nucleon suffered a hard collision, in all other interactions
with the target nucleons (calculated from geometry) it will behave as 
if nothing happened.  
There are some quantum mechanical (coherence) arguments to justify this
assumption, although it is quite remarkable to see them in the context
of defending a model whose basic tenet is the incoherent superposition
of independent, identical collisions.  

Other models take a different route to explain unexpected high-$p_T$
$p(d)$+A results.  In an early paper~\cite{levai2001} the authors
argue that {\it ``Our $p$A collision study showed that each $pp$ inelastic
collision adds $\sim$400\,MeV/$c$ transverse momentum to the partons
inside the proton (on average).  After a few such collisions the
partons gain high enough transverse momenta to become free of the
proton and during this transition time they do not interact (dead
time).  We assume that such a proton is ``lost'' for the reaction and
does not participate in particle production anymore.  We note that
such a picture correponds to a modification of the original Glauber
model.'' }

The color fluctuation model~\cite{alvioli2013} is another attempt to
explain the larger than expected $N_{coll}$ fluctuations in $p$+A
collisions.  The model, also called Glauber-Gribov model, allows the
total nucleon-nucleon cross-section $\sigma_{tot}$ to fluctuate and
thus change the distribution of wounded nucleons at a given impact
parameter, or, conversely, the impact parameter distribution for a
given number of wounded nucleons.  On the experimental side 
ATLAS~\cite{atlas2013-096} used two Glauber-Gribov parametrizations in
addition to the standard Glauber to describe their $E_T$ measurement
in $p$+Pb collisions.

\section{What next?}


We believe there is a potentially serious problem in establishing
collision geometry in very asymmetric collisions when at least one
hard scattering occurred, too.  We claim the existence of the problem
not because this or that particular measurement, odd or unexpected
result (like $R_{CP}$ for identified high $p_T$ particles), but
because of the experimental observation in Fig.~\ref{fig:ppbiasdata}
and the implausibility of the assumption that after a very hard
collision the projectile nucleon keeps interacting as an unexcited,
unchanged object.  For arguments' sake, let's take the opposite
extreme and assume that after a very hard scattering - which can turn
at any point of the eikonal traversing the target nucleus - the
projectile nucleon as a whole is simple ``out of the pool'', stops
interacting.  While this is obviously unrealistic 
({\it just as  no change whatsoever is}) 
it's easy to build a toy model around
it and check the experimental consequences.  

Our toy model is a standard Glauber Monte Carlo, with randomly
distributed, fixed size nucleons, the number of participants and
collisions are calculated with the hard disk approximation, and the
calculated soft production (charged multiplicity) is the
$N_{part}$-fold convolution of a realistic negative binomial
distribution (NBD).  The only significant difference w.r.t. the standard
Glauber model is that in each event we assign one of the collisions of
(exactly one) projectile nucleon as hard collision, and the higher the
$p_T$ generated, the more we reduce soft production by this nucleon
for the rest of its path in the target nucleus.  In other words, if in
the Glauber-picture the projectile nucleon scattered $n$ times, but
the $m$-th scattering was a hard one producing the maximum $p_T$ in
the event, then the total soft production will be calculated as an
$m$-fold convolution of the original NDB and an $n-m$-fold convolution
of a reduced NDB, where the mean of the reduced NBD decreases linearly
with increasing maximum $p_T$.
Since we are
interested in trends, namely, whether there are noticeable changes as
the maximum $p_T$ increases, we use a flat distribution for maximum
$p_T$ (instead of a realistic spectrum each $p_T$ is 
thrown with equal probability).

\begin{figure}
\centering{
\includegraphics[width=0.45\linewidth]{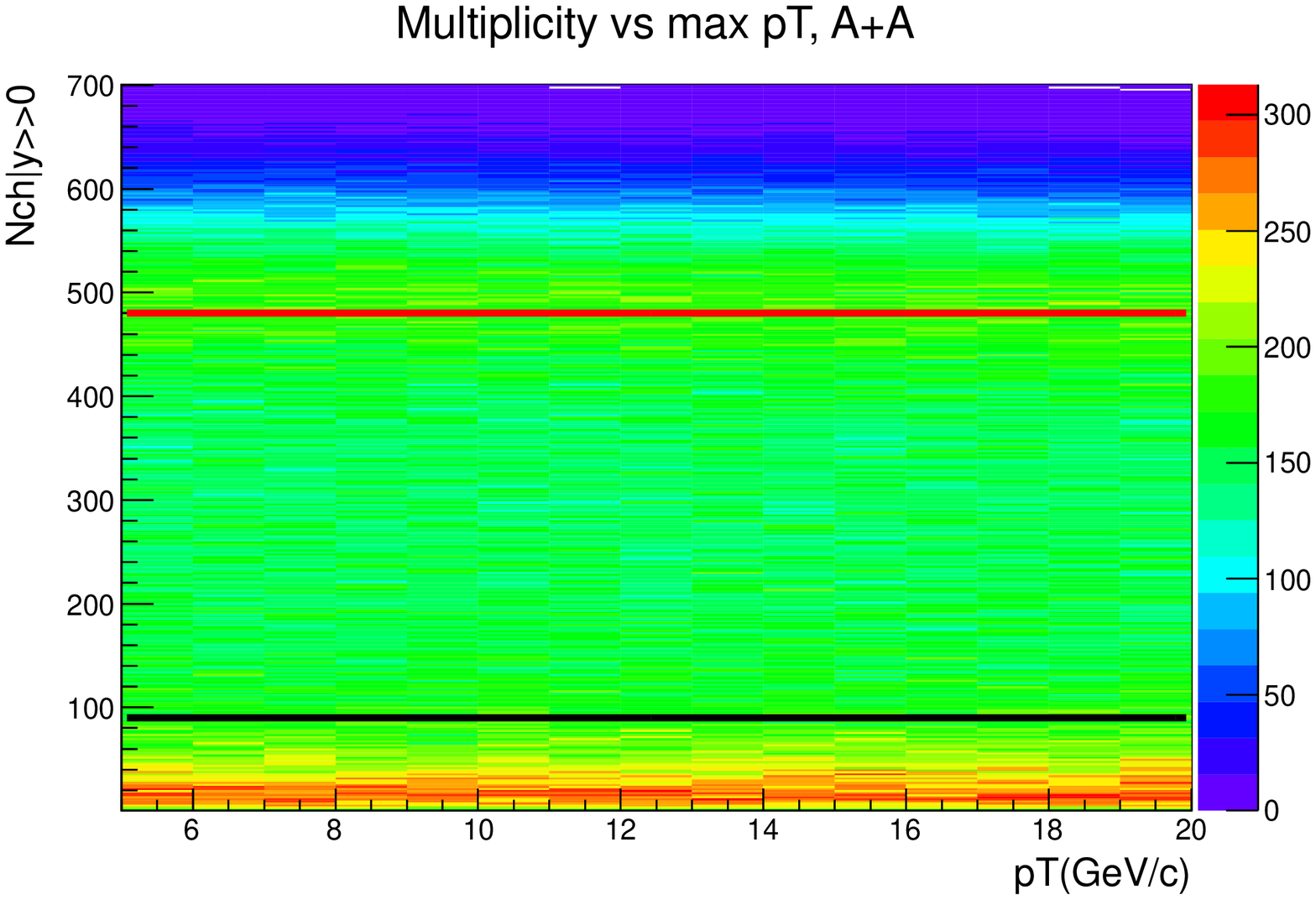}
\includegraphics[width=0.45\linewidth]{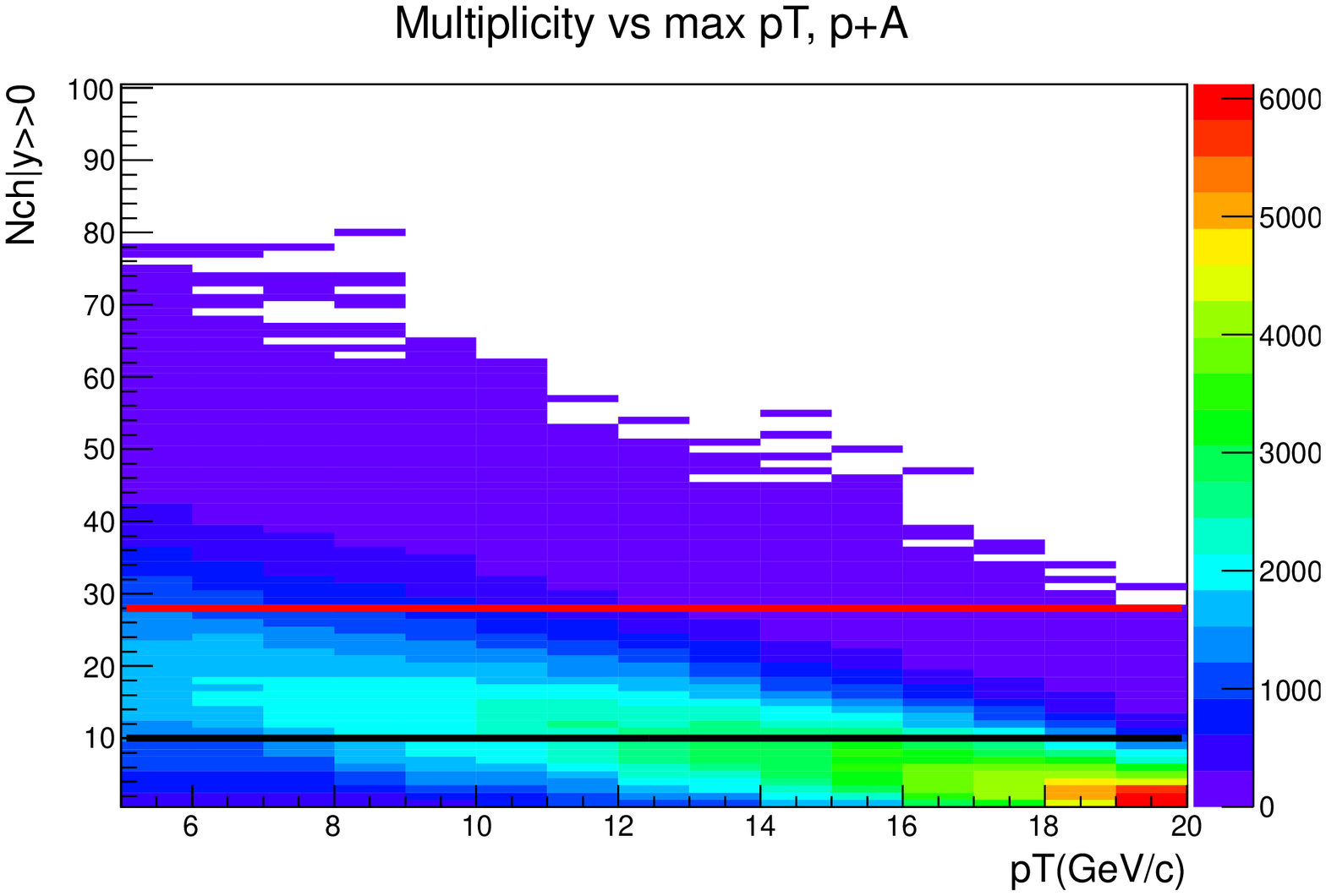}}
\caption{\label{fig:nchvspt}
  A simple model to illustrate possible differences between large-large
  and very small - large system collisions.  In both panels the
  vertical axis is a simulated soft multiplicity ($N^{TCD}_{ch}$) in
  the region where collision centrality is typically determined.  The
  horizontal axes are the maximum $p_T$ in the event; a hard collision
  producing high $p_T$ proportionally decreases further soft production
  from this participating nucleon.
  Left: large A+A system (here A=50).  Right: $p$+A system, with A=197.
  Red and black lines: limits for the 20\% most central and 20\% most
  peripheral collisions, based on the multiplicity distribution at low
  maximum $p_T$ (average event).  
}
\end{figure}

In Fig.~\ref{fig:nchvspt} we show the result of our toy model for two cases: 
first, when two large nuclei collide (left panel, A=50), second, when
a single proton collides with a large nucleus (right, $p$+Au).  As
already pointed out earlier, in large A+B systems, due to the large
number of participants in both nuclei ($N^A_{part}$, $N^B_{part}$) the
fact that one participant from each stops contributing to soft
particle production is irrelevant; the fluctuations ensure that the
multiplicity distribution - consequently, the multiplicity-based
centrality classes - are unchanged irrespective of the maximum $p_T$
observed in the event.  The situation is quite different for the
$p$+Au collision (right panel), where with increasing maximum $p_T$ 
a clear depletion of soft production is observed.  We believe such a
{\it triangular} shape with $p_T$ is a general property of very
asymmetric systems, while large-on-large systems essentially don't
change (they are ``rectangular'' with $p_T$). The red and black lines
indicate where the 20\% most central and 20\% most peripheral
collisions would be if assigned based on multiplicity in the average
(low $p_T$ only) events.  While there is no noticeable difference with
$p_T$ in A+A events, the character of the events in a centrality bin
changes a lot in asymmetric collisions.
While this toy model is deliberately simplistic, it describes features
seen in actual data quite well.

In summary, we reviewed briefly how fundamental geometry
like impact parameter or number of participants in heavy ion
collisions are connected to experimental observables.  We found that
as long as large colliding systems are
considered,  the correspondence between the theoretical quantities 
and experimental observables is quite unambigous even in the presence 
of a few very hard subprocesses.   However, in very asymmetric, 
specifically $p(d)$+A collisions the presence of a hard process
strongly biases the soft production, and, as a consequence, the
derived geometric quantities as well.


\ack
The author greatfully acknowledges inspiring discussions with
G. Barnafoldi, M. Gyulassy, P. Levai,  J. Nagle, N. Novitzky, 
T. Sakaguchi and M. J. Tannenbaum.




\end{document}